\documentclass{appolb}
\usepackage{epsfig}

\begin{document}
\pagestyle{plain}

\def\lsim{\raise0.3ex\hbox{$<$\kern-0.75em\raise-1.1ex\hbox{$\sim$}}}
\def\gsim{\raise0.3ex\hbox{$>$\kern-0.75em\raise-1.1ex\hbox{$\sim$}}}
\def\kk{{\bf k_1}}
\def\kg{{\bf k_2}}
\def\qq{{\bf q}}
\def\pom{{I\!\!P}}
\newcommand{\rk}{\mbox{\boldmath $k$}}
\newcommand{\rkp}{\mbox{\boldmath $k^{\prime}$}}
\newcommand{\rkn}{\mbox{$k$}}
\newcommand{\rr}{\mbox{\boldmath $r$}}
\newcommand{\rrn}{\mbox{$r$}}
\newcommand{\rp}{\mbox{\boldmath $p$}}
\newcommand{\rqq}{\mbox{\boldmath $q$}}
\renewcommand{\vec}[1]{\boldsymbol{#1}}
\newcommand{\dif}{\mathrm{d}}
\newcommand{\diff}[1]{\frac{\mathrm{d}#1}{#1}}
\newcommand{\xB}{x_{\scriptscriptstyle{B}}}
\newcommand{\chisq}{\chi^2/\mathrm{d.o.f.}}

\newcommand{\stt}{\small\tt}
\eqsec
\newcount\eLiNe\eLiNe=\inputlineno\advance\eLiNe by -1
\title{NUCLEAR DVCS WITHIN THE HIGH ENERGY QCD COLOR DIPOLE FORMALISM
\thanks{Email:{\tt magno.machado@unipampa.edu.br}}%
}
\author{Magno V. T. Machado
\address{Centro de Ci\^encias Exatas e Tecnol\'ogicas, Universidade Federal do Pampa \\
Campus de Bag\'e, Rua Carlos Barbosa. CEP 96400-970. Bag\'e, RS, Brazil}}
\maketitle

\begin{abstract}
In this contribution, we present a study of the coherent and incoherent nuclear DVCS process, $\gamma^* A \rightarrow \gamma\,X$, in the small-$x$ regime within the color dipole formalism. Predictions for the nuclear DVCS cross section at photon level in the collider kinematics are presented.
\end{abstract}

\section{Introduction}
An interesting way of probing hadronic matter involves the physics of deeply virtual Compton scattering (DVCS), where a parton in the proton absorbs the virtual photon, emits a real photon and the proton ground state is restored. At collider experiment, the relevant QCD diagrams involve the exchange of two gluons at low $x$ carrying different fractions of the initial proton momentum. The DVCS process thus measures generalized parton distributions (GPDs) which depends on two momentum fractions $x$ and $x^{\prime}$, as well as on $Q^2$ and the four-momentum transfer $t$ at the proton vertex. Similar process also occurs in $eA$ colliders, which is extremely sensitive to the corresponding nuclear parton distributions. Experimentally, DVCS on nucleons has been studied  by the H1 \cite{H1dvcs} and ZEUS \cite{ZEUSdvcs} Collaborations at DESY-HERA. Moreover, DVCS has been measured at low energies in CLAS  experiment \cite{CLAS} at the Jefferson Laboratory (JLAb) as well as in HERMES and COMPASS.  At high energies, nuclear DVCS can be studied on future electron-ion colliders (EICs).  For instance, the LHeC \cite{LHeC} project is a proposed colliding beam facility at CERN, which will exploit large energy and intensity provided by the LHC for lepton-nucleon (or nucleus) scattering. The large energy and the luminosity in such an experiment would allow the parton densities to be measured at unexplored momentum transfers $Q^2$ and small Bjorken $x\leq 10^{-6}$ for $Q^2\approx 1$ GeV$^2$. 

In this contribution, we present an alternative theoretical formalism for nuclear DVCS. Here, we present a summary of results obtained in Ref. \cite{MVTM}. We use the high energy color dipole approach \cite{dipole} to study the nuclear DVCS process at photon level. In order to do so, recent phenomenological models for the elementary dipole-hadron scattering amplitude that captures main features of the dependence on atomic number $A$, on energy and on momentum transfer $t$ are considered. This investigation is directly related and complementary to the conventional partonic description of nuclear DVCS, which considers the relevant nuclear GPDs. 

\section{DVCS on nucleons and nuclei in the color dipole approach}
Let us summarize the relevant formulas in the color dipole picture for the DVCS process on nucleons and nuclei. In such a framework \cite{dipole}, the scattering process $\gamma^* p\rightarrow \gamma p$ is assumed to proceed in three stages: first the incoming virtual photon fluctuates into a quark--antiquark pair, then the $q\bar{q}$ pair scatters elastically on the proton, and finally the $q\bar{q}$ pair recombines to form a real photon.  The imaginary part of the scattering amplitude for DVCS on nucleons is given by \cite{MPS,KMW,MW}
\begin{eqnarray}
 \mathcal{A}^{\gamma^* p\rightarrow \gamma p} =  \sum_f \sum_{h,\bar h} \int d^2r \int_0^1 dz\,\Psi^*_{h\bar h}(r,z,0)\,\mathcal{A}_{q\bar q}(x,r,\Delta)\,\Psi_{h\bar h}(r,z,Q)\,,
 \label{eq:ampvecm}
\end{eqnarray}
where $\Psi_{h\bar h}(r,z,Q)$ denotes the amplitude for a photon (with virtuality $Q$) to fluctuate into a quark--antiquark dipole with helicities $h$ and $\bar h$ and flavor $f$. The quantity $\mathcal{A}_{q\bar q}(x,r,\Delta)$ is the elementary amplitude for the scattering of a dipole of size $r$ on the proton, $\Delta $ denotes the transverse momentum lost by the outgoing proton (with $t=-\Delta^2$), $x$ is the Bjorken variable. As one has a real photon at the initial state, only the transversely polarized overlap function contributes to the cross section.  Summed over the quark helicities, for a given quark flavor $f$ it is given by \cite{MW},
\begin{eqnarray}
  (\Psi_{\gamma^*}^*\Psi_{\gamma})_{T}^f  =  \frac{N_c\,\alpha_{\mathrm{em}}e_f^2}{2\pi^2}\left\{\left[z^2+\bar{z}^2\right]\varepsilon_1 K_1(\varepsilon_1 r) \varepsilon_2 K_1(\varepsilon_2 r)  +   m_f^2 K_0(\varepsilon_1 r) K_0(\varepsilon_2 r)\right\},
  \label{eq:overlap_dvcs}
\end{eqnarray}
where we have defined the quantities $\varepsilon_{1,2}^2 = z\bar{z}\,Q_{1,2}^2+m_f^2$ and $\bar{z}=(1-z)$. Accordingly, the photon virtualities are $Q_1^2=Q^2$ (incoming virtual photon) and $Q_2^2=0$ (outgoing real photon). In what follows we set the quark masses as $m_{u,d,s}=0.14$ GeV for the light quarks and $m_c=1.4$ GeV for the charm quark. For the DVCS on nucleons, we take into account saturation models which successfully describe exclusive processes at high energies. In particular, we consider the non-forward saturation model of Ref. \cite{MPS} (hereafter MPS model), which captures the main features of the dependence on energy,  virtual photon virtuality and momentum transfer $t$. In the MPS model, the elementary elastic amplitude for dipole interaction is given by,
\begin{eqnarray}
\label{sigdipt}
\mathcal{A}_{q\bar q}(x,r,\Delta)= 2\pi R_p^2\,e^{-B|t|}N \left(rQ_{\mathrm{sat}}(x,|t|),x\right),
\end{eqnarray}
with the asymptotic behaviors $Q_{\mathrm{sat}}^2(x,\Delta)\sim
\max(Q_0^2,\Delta^2)\,\exp[-\lambda \ln(x)]$. Specifically, the $t$ dependence of the saturation scale is parametrised as
\begin{eqnarray}
\label{qsatt}
Q_{\mathrm{sat}}^2\,(x,|t|)=Q_0^2(1+c|t|)\:\left(\frac{1}{x}\right)^{\lambda}\,, \end{eqnarray}
in order to interpolate smoothly between the small and intermediate transfer
regions. The form factor $F(\Delta)=\exp(-B|t|)$ catches the transfer dependence of the proton vertex, which is factorised from the
projectile vertices and  does not spoil the geometric scaling properties. For the parameter $B$ we use the value $B=3.754$ GeV$^{-2}$ taken from Ref. \cite{MPS} (this parameter is reasonably stable in the phenomenological fits of MPS model). Finally, the scaling function $N$ appearing on Eq. (\ref{sigdipt}) is obtained from the forward saturation model \cite{Iancu:2003ge}. It has been shown in Ref. \cite{MVTM} that the MPS model describes all data on DVCS on nucleons at DESY-HERA energy.

\begin{figure}[t]
\centerline{\includegraphics[scale=0.4]{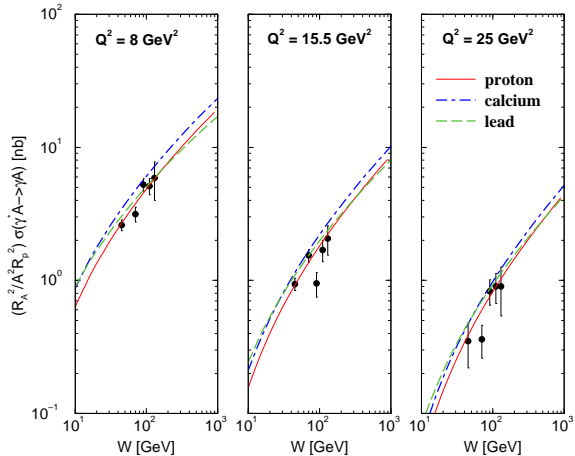}}
\caption{The coherent DVCS cross section rescaled by factor $A^2R_p^2/R_A^2$. The DVCS cross section on nucleons is also presented.}
\label{fig:1}
\end{figure}

Let us now present  a  study on the nuclear DVCS. In the situation when the recoiled nucleus is not detected, measurements of DVCS observables with nuclear targets involves the coherent and incoherent contributions. The coherent scattering corresponds to the case in which the nuclear target remains intact and it dominates at small $t$. The incoherent scattering occurs when the initial nucleus of atomic number $A$ transforms into the system of $(A-1)$ spectator bound/free nucleons and one interacting nucleon and it dominates at large $t$. We start looking at the coherent (elastic) nuclear DVCS contribution, $\gamma^* A \rightarrow \gamma A$, where the recoiled nucleus is intact. The implementation of nuclear effects in such a process is relatively   simple within the color dipole formalism in the low $x$ region. The master equation remains to be Eq. (\ref{eq:ampvecm}) replacing the elementary dipole-nucleon amplitude by an elementary dipole-nucleus amplitude. The usual procedure is to consider the Glauber-Gribov formalism for nuclear absorption. In phenomenological models in which geometric scaling is present (as in MPS saturation model) the extrapolation for a nucleus target is simplified. Based on the universatily bahavior for saturation models proposed in Ref. \cite{Armesto_scal}, in Ref. \cite{MVTM} we propose the following elementary elastic dipole-nucleus scattering amplitude,
\begin{eqnarray}
\label{sigdipnuc}
\mathcal{A}_{q\bar q}^{\mathrm{nuc}}(x,r,\Delta)= 2\pi R_A^2\,F_A(t)\,N \left(rQ_{\mathrm{sat},\,A};\,x\right),
\end{eqnarray}
where $F_A$ is the nuclear form factor and $Q_{\mathrm{sat},\,A}$ is the nuclear saturation scale (see Ref. \cite{MVTM} for details).

In diffractive incoherent (quasi-elastic) production of direct photons off nuclei, $\gamma^*A\rightarrow \gamma X$, one sums over all final states of the target nucleus except those which contain particle creation.  In order to compute the incoherent cross section we consider an approach involving the vector-dominance model (VDM) combined with the Glauber eikonal approximation. Explicit  calculations can be found in Ref. \cite{KopNik} in the approximation of a short coherence (or production) length, $\ell_c$, when one can treat the creation of the colorless $q\bar{q}$ pair as instantaneous compared to the formation length, $\ell_f$, which is comparable with the nuclear radius $R_A$. The expression for the incoherent cross section is given by \cite{KopNik}:
\begin{eqnarray}
\left. \frac{d\sigma^{T,L}}{dt}\right|_{t=0}=\int d^2b\,T_A(b)\left|\left\langle \Psi_{\gamma}^{T,L}\left|  \sigma_{dip}(x,r)\exp\left[-\frac{1}{2}\sigma_{dip}(x,r)T_A(b)\right]\right| \Psi_{\gamma^*}^{T,L} \right\rangle\right|^2,\nonumber \\
\label{incoh}
\end{eqnarray}
where $T_A(b)=\int_{-\infty}^{+\infty}dz\,\rho(b,z)$ is the nuclear thickness function given by the integral of the nuclear density along the trajectory at a given impact parameter $b$. The quantity $\sigma_{dip}$ is the forward dipole-target elastic amplitude, that is $\sigma_{dip}(x,r)=\mathcal{A}_{q\bar q}\,(x,r,\Delta = 0)$. The light-cone wavefunctions (and further integration on phase space) for transverse and longitudinal photons at initial and final state are labeled by $|\Psi_{\gamma,\,\gamma^*}^{T,L}\rangle$. The behavior on momentum transfer is slower in comparison to the coherent case and it is driven by the t-dependence of the cross section on quasi-free nucleons. In addition, it scales as $A$ in contrast to a $A^2$ scaling in the coherent case.

In Fig. \ref{fig:1} the coherent cross section is shown as a function of energy for distinct virtualities $Q^2=8,\,15.5,\,20$ GeV$^2$. For sake of comparison, the DVCS cross section on nucleons (solid lines) is also presented and compared with experimental measurements from DESY-HERA. We consider calcium (dot-dashed lines) and lead (dashed lines) nuclei \cite{MVTM}. The nuclear cross section has been rescaled by a factor $R_A^2/A^2R_p^2$ for illustration. The reason for that it is the  difficulty to compare the nuclear cross section to the Born term and the case $A=1$ does not match the DVCS cross section on proton (different $t$-slopes). It is verified a strong suppression for heavy nucleus at low $Q^2$. As the photon virtuality increases the corresponding suppression  diminishes. This fact is consistent with the general features of nuclear shadowing at small-$x$.

\begin{figure}[t]
\centerline{\includegraphics[scale=0.4]{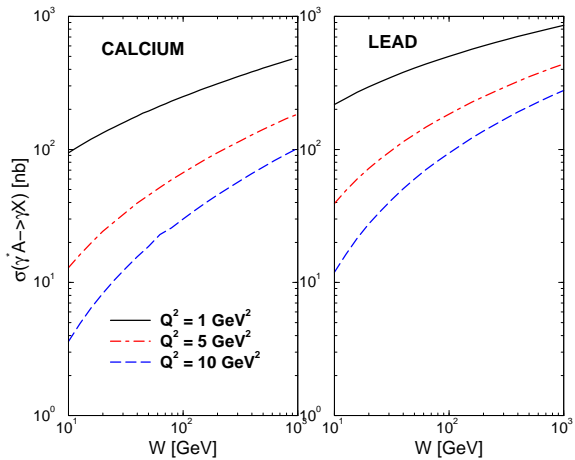}}
\caption{The incoherent DVCS cross section for calcium and lead nucleus as a function of energy for fixed values of photon virtualities.}
\label{fig:2}
\end{figure}

Finally, in Fig. \ref{fig:2} the incoherent DVCS cross section is presented. The estimation is done for calcium (left panel) and lead (right panel) nucleus as a function of energy for representative values of photon virtualities ($Q^2=1,\,5,\,10$ GeV$^2$. The incoherent is suppressed by a factor 3-4 for calcium and by factor 5-6 for lead in comparison to the coherent case \cite{MVTM}. This suppression can be understood from the different $A$-dependences of the integrated cross sections: the coherent DVCS cross section scales as $A^{4/3}$ whereas the incoherent cross section scales as $A$. Roughly speaking, the ratio incoherent/coherent scales as $A^{-1/3}$, which is consistent with the values found for the suppression. In order to illustrate the energy dependence of the incoherent DVCS cross section we parameterize it in the form $\sigma_{\mathrm{incoh}}=\sigma(W_0)[W/W_0]^{\alpha}$, with $W_0=100$ GeV.  For $Q^2=1$  GeV$^2$ one obtains $\sigma(W_0)=241,\,(480)$ nb, $\alpha=0.33,\,(0.27)$ for calcium (lead). It is verified that the effective energy exponent is a factor two smaller than the coherent case. This is directly associated to the strong exponential suppression for heavy nuclei appearing in Eq. (\ref{incoh}). That is, the dipole cross section attenuates with a constant absorption cross section.

As a summary, using the color dipole formalism we studied the DVCS process on nucleons and nuclei. Such an approach is robust in describing a wide class of exclusive processes measured at DESY-HERA and at the experiment CLAS (Jeferson Lab.), like meson production, diffractive DIS and DVCS. The theoretical uncertainties are smaller in this case in contrast to the exclusive vector meson production as the overlap photon function are well determined. We also provide estimations for the coherent and incoherent DVCS cross section, investigating their $A$-dependence. This is timely once DVCS off nuclei is a very promising tool for the investigation of the partonic structure of nuclei and it can be useful to clarify physics issues related to planned electron ion colliders (EIC's) as the LHeC.


\begin{thebibliography}{99}

\bibitem{H1dvcs}
  A.~Aktas {\it et al.}  [H1 Collaboration],
  Eur.\ Phys.\ J.\ C {\bf 44} (2005) 1; F.D. Aaron {\it et al.}  [H1 Collaboration], Phys.\ Lett.\ B {\bf 659} (2008) 796.

\bibitem{ZEUSdvcs}
  S.~Chekanov {\it et al.}  [ZEUS Collaboration],
  Phys.\ Lett.\ B {\bf 573} (2003) 46.

\bibitem{CLAS} S. Stepanyan {\it et al.}  [CLAS Collaboration], Phys. Rev. Lett. {\bf 87}, 182002, 2001.

\bibitem{LHeC} J.B. Dainton, M. Klein, P. Newman, E. Perez and F. Willeke, JINST {\bf 1}, P100001 (2006).

\bibitem{MVTM} M.V.T. Machado,  Eur.\ Phys.\ J.\  C {\bf 59}, 769 (2009).

\bibitem{dipole}  N.~N.~Nikolaev and B.~G.~Zakharov, Z. Phys. {\bf  C49}, 607
(1991); Z. Phys. {\bf C53}, 331 (1992); A.~H.~Mueller, Nucl. Phys.
{\bf B415}, 373 (1994); A.~H.~Mueller and B.~Patel, Nucl. Phys. {\bf B425}, 471
(1994).

\bibitem{KMW}
H. Kowalski, L. Motyka and G. Watt,
{\it Phys. Rev.} {\bf D74} (2006) 074016.


\bibitem{MPS}
C.~Marquet, R.~Peschanski and G.~Soyez,
  Phys.\ Rev.\  D {\bf 76}, 034011 (2007).

\bibitem{Watt} G. Watt and H. Kowalski, Phys. Rev. D {\bf 78}, 014016 (2008).

\bibitem{MW} L. Motyka, G. Watt, Phys. Rev. D {\bf 78}, 014023 (2008).

\bibitem{Guzey5} K. Goeke, V. Guzey and M. Siddikov, Eur.\ Phys.\ J.\ C {\bf 56}, 203 (2008).

\bibitem{MAPESO}
C. Marquet, R. Peschanski and G. Soyez,
{\it Nucl. Phys.} {\bf A756} (2005) 399.

\bibitem{Iancu:2003ge}
  E.~Iancu, K.~Itakura and S.~Munier,
  Phys.\ Lett.\ B {\bf 590} (2004) 199.

\bibitem{Shuvaev:1999ce}
A.~G.~Shuvaev {\it et al.},
Phys.\ Rev.\ D {\bf 60}, 014015 (1999).

\bibitem{Armesto_scal}
  N.~Armesto, C.~A.~Salgado and U.~A.~Wiedemann,
  Phys.\ Rev.\ Lett.\  {\bf 94}, 022002 (2005).

\bibitem{KopNik} B.Z. Kopeliovich and B.G. Zakharov, Phys. Rev. D {\bf 44}, 3466 (1991).

\end{thebibliography}
\end{document}